\documentclass[twocolumn]{jpsj3}
\usepackage{color}
\usepackage{bm}

\title{%
Superconducting Proximity Effect on a Two-Dimensional
Dirac Electron System
}

\author{%
Yositake {\sc Takane} and Ryo {\sc Ando}
}

\inst{%
Department of Quantum Matter, Graduate School of Advanced Sciences of Matter,\\
Hiroshima University, Higashihiroshima, Hiroshima 739-8530, Japan
}

\recdate{ \hspace{50mm} }

\abst{%
The superconducting proximity effect on two-dimensional massless 
Dirac electrons is usually analyzed using a simple model consisting of
the Dirac Hamiltonian and an energy-independent pair potential.
Although this conventional model is plausible,
it is questionable whether it can fully describe
the proximity effect from a superconductor.
Here, we derive a more general proximity model
starting from an appropriate microscopic model
for the Dirac electron system in planar contact with a superconductor.
The resulting model describes the proximity effect in terms of
the energy-dependent pair potential and renormalization term.
Within this framework,
we analyze the density of states, the quasiparticle wave function,
and the charge conservation of Dirac electrons.
The result reveals several characteristic features of the proximity effect,
which cannot be captured within the conventional model.
}

\kword{%
graphene, strong topological insulator, Dirac equation, superconductor
}

\begin{document}
\sloppy
\maketitle

\section{Introduction}

Since the isolation of monolayer graphene,~\cite{novoselov}
two-dimensional (2D) massless Dirac electron systems have attracted
considerable attention in the condensed matter community.
This attention has been intensified by the discovery of 2D Dirac electrons
on the surface of strong topological insulators.~\cite{fu1,moore,roy}
Low-energy electrons in monolayer graphene are referred to as Dirac electrons
as they obey the massless Dirac equation.~\cite{wallace}
A three-dimensional strong topological insulator is insulating
in the bulk but hosts metallic electron states on its surface.
The surface electrons of strong topological insulators are confined
in a thin surface region and obey the massless Dirac equation,~\cite{liu}
so they are also regarded as 2D Dirac electrons.

In this paper, we focus on the proximity effect on 2D Dirac electrons
coupled with a bulk superconductor.
Such a setup is most naturally created by
depositing a superconductor on top of a graphene sheet~\cite{heersche,du,sato}
or a topological insulator.~\cite{sacepe,zhang}
Consequently, the resulting hybrid system has a 2D planar structure.
The simplest way to describe electron states in the region covered by
a superconductor is to add an effective pair potential $\Delta_{\rm eff}$
to the Hamiltonian of the Dirac electron system.~\cite{volkov}
Let us consider the 2D Dirac electron system on the $xy$ plane
governed by the following $2 \times 2$ Dirac Hamiltonian:
\begin{align}
       \label{eq:def-DH}
   \check{H}_{\rm D}^{0}
   = \left[
       \begin{array}{cc}
         -\mu & v\hat{k}_{-} \\
         v\hat{k}_{+} & -\mu
       \end{array}
     \right] ,
\end{align}
where $v$ and $\mu$ are the velocity and chemical potential, respectively,
and $\hat{k}_{\pm}\equiv -{\rm i}\partial_{x}\pm \partial_{y}$.
To theoretically treat quasiparticle states under the proximity effect,
we need to use the Bogoliubov-de Gennes (BdG) equation,~\cite{de-gennes}
which can describe the mixing of electron and hole states
induced by an effective pair potential.
The corresponding $4 \times 4$ Hamiltonian for the covered region becomes
\begin{align}
      \label{eq:def-BdG_0}
   H_{\rm BdD}^{0}
   = \left[
       \begin{array}{cc}
         \check{H}_{\rm D}^{0} & \check{\Delta} \\
         \check{\Delta} & -\check{H}_{\rm D}^{0}
       \end{array}
     \right] ,
\end{align}
where
\begin{align}
   \check{\Delta}
   = \left[
       \begin{array}{cc}
         \Delta_{\rm eff} & 0 \\
         0 & \Delta_{\rm eff}
       \end{array}
     \right] .
\end{align}
The model presented above was first proposed for graphene
by Beenakker,~\cite{beenakker}
and has been widely applied to not only superconductor-graphene
junctions~\cite{titov,bhattacharjee,moghaddam,linder} but also
superconductor-topological insulator junctions.~\cite{fu2,akhmerov,tanaka}
Hereafter, it is referred to as the conventional model.

Although the conventional model is plausible,
its microscopic justification has been lacking in a strict sense.
Note that $H_{\rm BdD}^{0}$ cannot be distinguished from
the Hamiltonian describing Dirac electrons in the superconducting state.
Furthermore, it is difficult to apply the conventional model to the analysis
of the temperature ($T$) dependence of physical quantities
because $\Delta_{\rm eff}$ is a phenomenological parameter
and its $T$-dependence is not easy to determine.
It is thus questionable whether the proximity effect is
fully described by this model.

The purpose of this paper is to establish a more general proximity model.
Starting from an appropriate microscopic model
for the 2D Dirac electron system in planar contact with a bulk superconductor,
we derive an effective model for Dirac electrons by exactly integrating out
the electron degrees of freedom in the superconductor.
In the resulting effective model, the proximity effect is represented by
the energy-dependent pair potential $\phi$ and renormalization term $\eta$.
In the Matsubara representation
with $\omega$ being the fermion Matsubara frequency,
the effective Hamiltonian is given by
\begin{align}
      \label{eq:H_BdG_omega}
  & H_{\rm BdG}(\omega) =
     \nonumber \\
  & \left[
      \begin{array}{cccc}
        -\mu-\eta(\omega) & v\hat{k}_{-} & \phi(\omega) & 0 \\
        v\hat{k}_{+} & -\mu-\eta(\omega) & 0 & \phi(\omega) \\
        \phi(\omega) & 0 & \mu-\eta(\omega) & -v\hat{k}_{-} \\
        0 & \phi(\omega) & -v\hat{k}_{+} & \mu-\eta(\omega)
      \end{array}
    \right]
\end{align}
with
\begin{align}
      \label{eq:def-eta}
   \eta(\omega) & = \frac{{\rm i}\Gamma\omega}{\Omega(\omega)} ,
          \\
      \label{eq:def-phi}
   \phi(\omega) & = \frac{\Gamma\Delta_{0}}{\Omega(\omega)} ,
\end{align}
where $\Delta_{0}$ is the pair potential of
the bulk superconductor,~\cite{comment0}
$\Gamma$ characterizes the coupling strength of
Dirac electrons to the superconductor, and
\begin{align}
  \Omega(\omega) = \sqrt{\omega^{2}+\Delta_{0}^{2}} .
\end{align}
The effective Hamiltonian with real energy $\epsilon$
is simply obtained by carrying out the analytic continuation of
${\rm i}\omega \to \epsilon + {\rm i}\delta$,
where $\delta$ is a positive infinitesimal.
Within this framework, we analyze the density of states,
the quasiparticle wave function, and the charge conservation
of Dirac electrons
to reveal the characteristic features of the proximity effect.

Here, it is fair to mention that Sau {\it et al}.~\cite{sau} have
derived a model essentially equivalent to Eq.~(\ref{eq:H_BdG_omega})
by adapting the argument of McMillan~\cite{mcmillan}
to a microscopic model for a superconductor-topological insulator junction.
However, they focused on the low-frequency limit of $\omega \to 0$
and did not explicitly examine the role of
the $\omega$-dependence of the effective Hamiltonian.
Furthermore, the derivation of Eq.~(\ref{eq:H_BdG_omega}) given below
is simpler and more transparent than that in Ref.~\citen{sau}.
The same model was also proposed in
Refs.~\citen{takane1} and \citen{takane2}
for a superconductor-graphene junction.

Our other task in this study is to determine the relationship
between $H_{\rm BdG}(\omega)$ and $H_{\rm BdG}^{0}$.
In Ref.~\citen{takane2}, an expression for the Josephson current
through the Dirac electron system is derived
on the basis of $H_{\rm BdG}(\omega)$.
It is shown that, at $T = 0$,
the expression in the strong coupling limit of $\Gamma \gg \Delta_{0}$
reproduces that of Titov and Beenakker derived on the basis of
$H_{\rm BdD}^{0}$ with $\Delta_{\rm eff} = \Delta_{0}$,~\cite{comment1}
suggesting some relationship between the two Hamiltonians.
We show that, in spite of the apparent difference between them,
the behavior of quasiparticles described by $H_{\rm BdG}(\omega)$
becomes nearly identical to that described by $H_{\rm BdG}^{0}$
under the condition of $\mu \gg \Gamma \gg \Delta_{0}$.
This accounts for the correspondence of the two expressions for
the Josephson current in the strong coupling limit.

In the next section, we introduce a simple microscopic model
for the Dirac electron system in planar contact with a bulk superconductor,
and derive the effective Hamiltonian $H_{\rm BdG}$ for Dirac electrons
fully taking account of the proximity effect from a superconductor.
In Sect. 3, we calculate the density of states in the Dirac electron system
as a simple application of $H_{\rm BdG}$.
In Sect. 4, we obtain the wave function of quasiparticle states
by solving the BdG equation for $H_{\rm BdG}$.
It is shown that the resulting wave function is nearly identical to
that obtained from the conventional model Hamiltonian $H_{\rm BdG}^{0}$
when $\mu \gg \Gamma \gg \Delta_{0}$,
indicating that $H_{\rm BdG}$ and $H_{\rm BdG}^{0}$ describe
the same physics under this condition.
The charge conservation in the system described by $H_{\rm BdG}$
is considered in Sect. 5.
The last section is devoted to a summary.
We set $\hbar = k_{\rm B} = 1$ throughout this paper.

\section{Model and Formulation}

\begin{figure}[btp]
\begin{center}
\includegraphics[height=3.0cm]{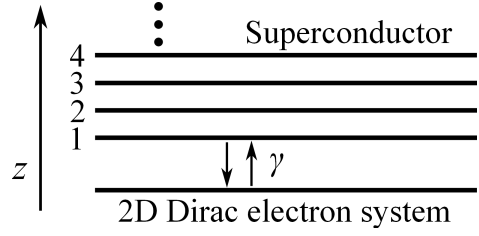}
\end{center}
\caption{2D Dirac electron system in planar contact
with a layered superconductor.
}
\end{figure}
Let us consider the 2D Dirac electron system on the $xy$ plane
in planar contact with a bulk superconductor.
For convenience of analysis, we assume that the superconductor
consists of an infinite number of 2D superconducting layers stacked
in the $z$-direction, and that the first layer is coupled with
the Dirac electron system (see Fig.~1).
We assume that the system is translationally invariant in the $x$-
and $y$-directions, and treat the in-plane wave vector
$\mib{k}_{\parallel} \equiv (k_{x},k_{y})$ as a conserved quantity.

The action $S$ for this system is decomposed into
$S=S_{\rm D}+S_{\rm S}+S_{\rm T}$, where $S_{\rm D}$ and $S_{\rm S}$
respectively describe the 2D Dirac electron system and
the bulk superconductor,
and $S_{\rm T}$ corresponds to the coupling between them.
We consider only the $\mib{k}_{\|}$-component
(and its time-reversed partner) of $S$ in the following argument
since $\mib{k}_{\|}$ is a good quantum number,
and use the Matsubara representation.
Firstly, we present an expression for $S_{\rm S}$.
Let $\psi_{j\sigma}$ be the electron field with spin $\sigma$
for the $j$th layer of the superconductor ($j = 1,2,3,\dots$).
Here and hereafter, we do not explicitly indicate
the $\mib{k}_{\|}$-dependence of the electron field.
The adjacent layers are coupled by the transfer integral $t$,
and the in-plane energy dispersion in each layer is
$\epsilon(\mib{k}_{\|})\equiv \mib{k}_{\|}^{2}/(2m)$ in the normal state.
We assume that the pair potential $\Delta_{0}$
and the chemical potential $\mu_{\rm S}$ are constant over all layers.
Taking all these into account, $S_{\rm S}$ is given by
\begin{align}
       \label{action_S}
   S_{\rm S}
 & = T\sum_{\omega} \sum_{j\ge 1}
     \Bigg[ \sum_{\sigma=\uparrow,\downarrow}
             \Big(
               \psi_{j\sigma}^{\dagger}(\omega)
               \left(-{\rm i}\omega-\mu_{\rm S}(\mib{k}_{\parallel}) \right)
               \psi_{j\sigma}(\omega)
    \nonumber \\
 & \hspace{10mm}
              -t \psi_{j\sigma}^{\dagger}(\omega)\psi_{j+1\sigma}(\omega)
              -t \psi_{j+1\sigma}^{\dagger}(\omega)\psi_{j\sigma}(\omega)
             \Big)
    \nonumber \\
 & \hspace{10mm}
     + \Delta_{0} \psi_{j\uparrow}^{\dagger}(\omega)
             \psi_{j\downarrow}^{\dagger}(-\omega)
     + \Delta_{0} \psi_{j\downarrow}(-\omega)\psi_{j\uparrow}(\omega)
    \Bigg] ,
\end{align}
where $\mu_{\rm S}(\mib{k}_{\|})$ is the effective chemical potential
defined by
$\mu_{\rm S}(\mib{k}_{\|}) = \mu_{\rm S}-\epsilon(\mib{k}_{\|})$.
Next we present an expression for $S_{\rm D}$.
We assume that the Dirac electron system is simply described by
the Hamiltonian (\ref{eq:def-DH}).
Let $\psi_{\rm D\sigma}$ be the Dirac electron field with spin $\sigma$.
To treat the superconducting proximity effect,
we must simultaneously consider electrons and holes.~\cite{comment2}
This is achieved by employing the four-component field
$\Psi_{\rm D}(\omega)$:
\begin{align}
\Psi_{\rm D}(\omega)
\equiv \,\raisebox{3.6mm}{}^{\rm t}\!
\left(\psi_{\rm D\uparrow}(\omega),
\psi_{\rm D\downarrow}(\omega),
\psi_{\rm D\downarrow}^{\dagger}(-\omega),
-\psi_{\rm D\uparrow}^{\dagger}(-\omega)\right) .
\end{align}
The corresponding action $S_{\rm D}$ is written as
\begin{align}
       \label{action_D}
   S_{\rm D}
   = T\sum_{\omega} \frac{1}{2}
     \Psi_{\rm D}^{\dagger}(\omega)
     \left(-{\rm i}\omega 1_{4\times4}+ H_{\rm D} \right)\Psi_{\rm D}(\omega) ,
\end{align}
where $1_{4\times4}={\rm diag}\{1,1,1,1\}$,
\begin{align}
   H_{\rm D}
   = \left[
       \begin{array}{cc}
         \check{H}_{\rm D}^{0} & \check{0} \\
         \check{0} & -\check{H}_{\rm D}^{0}
       \end{array}
     \right] ,
\end{align}
and the factor $1/2$ is attached to avoid double counting.
Finally, the coupling term is expressed as
\begin{align}
       \label{action_S}
   S_{\rm T}
   = T\sum_{\omega}\sum_{\sigma=\uparrow,\downarrow}
     \Big[ -\gamma \psi_{1\sigma}^{\dagger}(\omega)
                   \psi_{\rm D\sigma}(\omega)
           -\gamma \psi_{\rm D\sigma}^{\dagger}(\omega)
                   \psi_{1\sigma}(\omega)
     \Big] ,
\end{align}
where $\gamma$ is the transfer integral connecting
the Dirac electron system and the first layer of the bulk superconductor.

The proximity correction $S_{\Sigma}$ to the Dirac electron system
is expressed as
\begin{align}
     \label{eq:def-Sigma}
  \exp\left(-S_{\Sigma}\right)
  = \frac{\int \prod_{j,\sigma,\omega}
          D\psi_{j\sigma}(\omega)D\psi_{j\sigma}^{\dagger}(\omega)
          \exp\left(-S_{\rm S}-S_{\rm T}\right)}
         {\int \prod_{j,\sigma,\omega}
          D\psi_{j\sigma}(\omega)D\psi_{j\sigma}^{\dagger}(\omega)
          \exp\left(-S_{\rm S}\right)} .
\end{align}
Hence, the effective action for the Dirac electron system is given by
$S_{\rm eff} \equiv S_{\rm D}+S_{\Sigma}$.
We obtain an expression for $S_{\Sigma}$ by integrating out
$\psi_{j\sigma}$ and $\psi_{j\sigma}^{\dagger}$ for all $j$.~\cite{affleck}
The derivation of $S_{\Sigma}$ is presented in Appendix A.
The result is
\begin{align}
     \label{eq:Sigma_prox}
   S_{\Sigma}
 & = T\sum_{\omega}\frac{1}{2}
     \Psi_{\rm D}^{\dagger}(\omega)
     \left(
       \begin{array}{cc}
         V(\omega)\check{1} & \chi\phi(\omega)\check{1} \\
         \chi\phi(\omega)\check{1} & -V(-\omega)\check{1}
       \end{array}
     \right)
     \Psi_{\rm D}(\omega) ,
\end{align}
where $\check{1}={\rm diag}\{1,1\}$, $\phi(\omega)$ is defined
in Eq.~(\ref{eq:def-phi}), and
\begin{align}
  V(\omega) = \frac{\Gamma\mu_{\rm S}(\mib{k}_{\parallel})}{2t}
                -\chi\frac{{\rm i}\Gamma\omega}{\Omega(\omega)}
\end{align}
with
\begin{align}
    \label{eq:def-chi}
   \chi
   = 1-\frac{1}{2}\left(\frac{\mu_{\rm S}(\mib{k}_{\|})}{2t}
                  \right)^{2} .
\end{align}
Here, the coupling strength $\Gamma$ is defined by
\begin{align}
  \Gamma = \frac{\gamma^{2}}{t} .
\end{align}
As $2t$ is the largest energy scale of the system under consideration,
we can safely ignore the first term of $V(\omega)$ and set $\chi = 1$.
With this reduction, the effective action is simplified to
\begin{align}
   S_{\rm eff}
 & = T\sum_{\omega}\frac{1}{2}
     \Psi_{\rm D}^{\dagger}(\omega)
     \big(-{\rm i}\omega 1_{4\times 4} +H_{\rm BdG}(\omega) \big)
     \Psi_{\rm D}(\omega) ,
\end{align}
where $H_{\rm BdG}$ is the $\omega$-dependent effective Hamiltonian
presented in Eq.~(\ref{eq:H_BdG_omega}).
That is, Dirac electrons under the proximity effect
are described by this effective Hamiltonian.
It should be emphasized that, in its derivation,
we do not rely on a perturbative treatment with respect to
the coupling term $S_{\rm T}$ (see Appendix A).
Hence, this approach is applicable even when the coupling is very strong.

In the above argument, we properly take account of
the proximity effect on the Dirac electron system
but completely ignore the reverse effect on the superconductor.
Generally, the strength of the proximity effect
is determined by the density of states in the partner
to which the system under consideration is coupled.~\cite{mcmillan}
As the density of states of Dirac electrons is much smaller than
that of the bulk superconductor,
the reverse effect can be safely ignored.

\section{Density of States}

As a simple application of the effective Hamiltonian $H_{\rm BdG}(\omega)$
derived in the previous section, we analyze the density of states
in the Dirac electron system under the proximity effect.
The detailed behavior of quasiparticle states is analyzed in the next section.

Let us introduce the thermal Green's function defined as
\begin{align}
 \mathcal{G}(\mib{k}_{\|},\omega)
 = \left[{\rm i}\omega 1_{4\times 4}
         -H_{\rm BdG}(\omega)\big|_{\hat{k}_{\pm}\to k_{x}\pm{\rm i}k_{y}}
   \right]^{-1} ,
\end{align}
in terms of which the density of states $D(\epsilon)$
at energy $\epsilon$ is expressed as
\begin{align}
 D(\epsilon)
   = -\frac{1}{\pi}\int\frac{{\rm d}k_{\|}^2}{(2\pi)^{2}}
     {\rm Im} \left\{ {\rm tr}\left\{{\mathcal G}(\mib{k}_{\|},\omega)
                              \right\}
                      \big|_{{\rm i}\omega \to \epsilon + {\rm i}\delta}
              \right\} ,
\end{align}
where $\delta$ is a positive infinitesimal.
It is easy to show that
\begin{align}
  {\rm tr}\left\{{\mathcal G}(\mib{k}_{\|},\omega)\right\}
 & = \frac{-2{\rm i}\tilde{\omega}}
          {(vk_{\|}-\mu)^2+\phi(\omega)^2+\tilde{\omega}^2}
       \nonumber \\
 &   \hspace{2mm}
   + \frac{-2{\rm i}\tilde{\omega}}
          {(vk_{\|}+\mu)^2+\phi(\omega)^2+\tilde{\omega}^2} ,
\end{align}
where $\tilde{\omega}=(1+\Gamma/\Omega)\omega$ and $k_{\|}=|\mib{k}_{\|}|$.
To avoid the unphysical divergence of the integral over $k_{\|}$,
we introduce the cutoff energy $\epsilon_{\rm c}$ that is equivalent
to half of the band width.
After analytic continuation,
the energy-dependent pair potential becomes
\begin{align}
  \phi(\epsilon)
  = \frac{\Gamma\Delta_{0}}{\sqrt{\Delta_{0}^{2}-\epsilon_{+}^{2}}} ,
\end{align}
where
\begin{align}
  \epsilon_{+}
  = \epsilon + {\rm i}\delta .
\end{align}
Note that the behavior of $\phi(\epsilon)$ markedly changes depending on
whether $\epsilon$ is greater or smaller than $\Delta_{0}$ as follows:
\begin{align}
  \phi(\epsilon)
  = \left\{ \begin{array}{cc}
              \frac{\Gamma\Delta_{0}}{\sqrt{\Delta_{0}^{2}-\epsilon^{2}}}
                & (\Delta_{0} > \epsilon \ge 0) \\
              {\rm i}\frac{\Gamma\Delta_{0}}
              {\sqrt{\epsilon^{2}-\Delta_{0}^{2}}}
                & (\epsilon > \Delta_{0}) .
            \end{array}
    \right.
\end{align}
It is convenient to introduce the renormalization factor
\begin{align}
   Z(\epsilon)
   = 1+\frac{\Gamma}{\sqrt{\Delta_{0}^{2}-\epsilon_{+}^{2}}}
\end{align}
and the effective pair potential
\begin{align}
      \label{eq:def-Del_eff}
   \Delta(\epsilon)
   = \frac{\phi(\epsilon)}{Z(\epsilon)} .
\end{align}
As we see below, this effective pair potential determines
the proximity-induced energy gap of Dirac electrons.
Obviously, the $\epsilon$-dependence of the pair potential
is completely ignored in the conventional model.

Carrying out the integration over $k_{\|}$, we finally obtain
\begin{align}
    \label{eq:Dos_gen}
 D(\epsilon)
 & = \frac{1}{\pi^{2}v^{2}}
     {\rm Im}
     \Bigg[ \epsilon Z(\epsilon)
            \ln\left(\frac{\epsilon_{\rm c}^{2}-\mu^{2}}
                          {Z(\epsilon)^2\Theta(\epsilon)^{2}+\mu^{2}}
               \right)
      \nonumber \\
 & \hspace{18mm}
          + \frac{{\rm i}\mu\epsilon}{\Theta(\epsilon)}
            \ln\left(\frac{{\rm i}Z(\epsilon)\Theta(\epsilon)+\mu}
                          {{\rm i}Z(\epsilon)\Theta(\epsilon)-\mu}
               \right)
     \Bigg] ,
\end{align}
where
\begin{align}
       \label{eq:def-Omega_epsi}
   \Theta(\epsilon)
   = \sqrt{\Delta(\epsilon)^2-\epsilon_{+}^{2}} .
\end{align}
As $Z(\epsilon)$ and hence $\Delta(\epsilon)$ are real numbers
when $\epsilon < \Delta_{0}$, it is easy to see that
\begin{align}
  \Theta(\epsilon)
  = \left\{ \begin{array}{cc}
              \sqrt{\Delta(\epsilon)^2-\epsilon^{2}}
                & (\Delta(\epsilon) > \epsilon \ge 0) \\
              -{\rm i}\sqrt{\epsilon^{2}-\Delta(\epsilon)^2}
                & (\Delta_{0} > \epsilon > \Delta(\epsilon)) .
            \end{array}
    \right.
\end{align}
Accordingly, the density of states vanishes
when $\Delta(\epsilon) > \epsilon \ge 0$ because the function
in the square brackets of Eq.~(\ref{eq:Dos_gen}) has no imaginary part.
This indicates that the proximity-induced energy gap $\epsilon_{\rm g}$
in the Dirac electron system is determined by
\begin{align}
       \label{eq:def-gap}
  \epsilon_{\rm g} = \Delta(\epsilon_{\rm g}) .
\end{align}
It is easy to show that $\epsilon_{\rm g}$ in the weak coupling limit
of $\Delta_{0} \gg \Gamma$ is approximated as
\begin{align}
     \label{eq:gap_1}
  \epsilon_{\rm g} = \frac{\Delta_{0}\Gamma}{\Delta_{0}+\Gamma} ,
\end{align}
while, in the opposite strong coupling limit, we find
\begin{align}
     \label{eq:gap_2}
  \epsilon_{\rm g} = \Delta_{0} - \frac{2\Delta_{0}^{3}}{\Gamma^{2}} .
\end{align}
Equation~(\ref{eq:gap_2}) indicates that the energy gap approaches
$\Delta_{0}$ in the limit of $\Gamma/\Delta_{0} \to \infty$.
The energy gap numerically determined as a function of $\Gamma/\Delta_{0}$
is shown in Fig.~2, where the dotted (dashed) line represents
the approximate expression for the weak coupling (strong coupling) limit.
\begin{figure}[btp]
\begin{center}
\includegraphics[height=4.5cm]{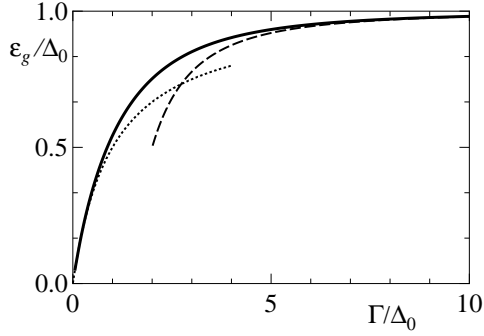}
\end{center}
\caption{Energy gap $\epsilon_{\rm g}$ as a function of $\Gamma/\Delta_{0}$.
}
\end{figure}

In the region of $\Delta_{0} > \epsilon > \epsilon_{\rm g}$,
we can easily extract the imaginary part of the function
in the square brackets and simply express the density of states as
\begin{align}
    \label{eq:DOS_red}
 D(\epsilon)
 & = \frac{\epsilon}{\pi v^{2}}
     \Bigg[ Z(\epsilon)
            \theta\left(Z(\epsilon)\sqrt{\epsilon^{2}-\Delta(\epsilon)^{2}}
                        -\mu
                  \right)
        \nonumber \\
 & \hspace{12mm}
          + \frac{\mu}{\sqrt{\epsilon^{2}-\Delta(\epsilon)^{2}}}
            \theta\left(\mu-
                        Z(\epsilon)\sqrt{\epsilon^{2}-\Delta(\epsilon)^{2}}
                  \right)
   \Bigg] ,
\end{align}
where $\theta(x)$ is the Heaviside step function.
It is interesting to note that the familiar square-root singularity
[i.e., the second term of Eq.~(\ref{eq:DOS_red})]
disappears when $\mu \approx 0$.
This should be regarded as a characteristic feature of
the Dirac electron system.

\begin{figure}[btp]
\begin{center}
\includegraphics[height=5.5cm]{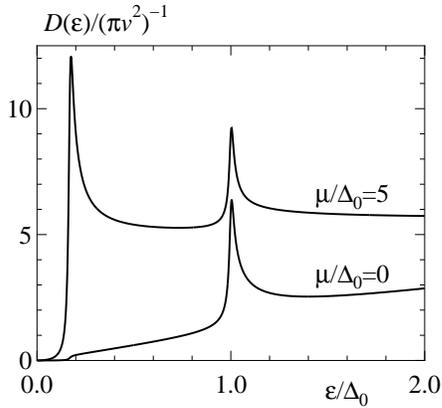}
\end{center}
\caption{Density of states in the Dirac electron system
in the cases of $\mu/\Delta_{0}=0$ and $5$
with $\Gamma/\Delta_{0}=0.2$.
}
\end{figure}
The density of states for arbitrary $\epsilon$
is numerically obtained from Eq.~(\ref{eq:Dos_gen}).
As an example, we calculate $D(\epsilon)$ at $\Gamma/\Delta_{0} = 0.2$
in the cases of $\mu/\Delta_{0} = 0$ and $5$
with $\epsilon_{\rm c}/\Delta_{0}=2000$.
To simulate inelastic scattering effects,
which are inevitably present in actual experimental situations,
we set the positive infinitesimal $\delta$
to be $\delta/\Delta_{0} = 0.01$.
The result is shown in Fig.~3.
The peak structure at $\epsilon/\Delta_{0} = 1$ reflects
the square-root singularity of the density of states
in the bulk superconductor.
Obviously, this cannot be captured within the conventional model
with the energy-independent pair potential $\Delta_{\rm eff}$
since information on the bulk superconductor
is not adequately encoded in it.
We observe that, at the gap edge (i.e., $\epsilon = \epsilon_{\rm g}$),
there is no singular behavior in the case of $\mu/\Delta_{0}=0$,
while a sharp enhancement appears in the case of $\mu/\Delta_{0}=5$,
in accordance with Eq.~(\ref{eq:DOS_red}).

\section{Quasiparticle States}

We introduce an effective BdG equation with real energy $\epsilon$
and obtain its exact eigenstates.
Considering the resulting eigenstate wave function,
we demonstrate that our model becomes essentially equivalent to
the conventional model in a certain limit.

The effective Hamiltonian for the BdG equation is obtained
by carrying out the analytic continuation
${\rm i}\omega \to \epsilon + {\rm i}\delta$ in $H_{\rm BdG}(\omega)$.
Then the BdG equation is given in the following form:
\begin{align}
     \label{eq:BdG}
  H_{\rm BdG}(\epsilon)\Psi(x,y) = \epsilon\Psi(x,y) ,
\end{align}
where
\begin{align}
     \label{eq:H_real-energy}
  & H_{\rm BdG}(\epsilon) =
       \nonumber \\
  & \left[
      \begin{array}{cccc}
        -\mu-\eta(\epsilon) & v\hat{k}_{-} & \phi(\epsilon) & 0 \\
        v\hat{k}_{+} & -\mu-\eta(\epsilon) & 0 & \phi(\epsilon) \\
        \phi(\epsilon) & 0 & \mu-\eta(\epsilon) & -v\hat{k}_{-} \\
        0 & \phi(\epsilon) & -v\hat{k}_{+} & \mu-\eta(\epsilon)
      \end{array}
    \right]
\end{align}
with $\eta(\epsilon) = \left(Z(\epsilon)-1\right)\epsilon$.

We hereafter assume that $\Psi(x,y)$ varies as
${\rm e}^{{\rm i}k_{y}y}$ in the $y$ direction with a real $k_{y}$,
and hence $\Psi(x,y)$ is rewritten as
\begin{align}
  \Psi(x,y)\equiv{\rm e}^{{\rm i}k_{y}y}\Psi(x) .
\end{align}
Note that the BdG equation has evanescent solutions.
That is, $\Psi(x)$ can be
an exponentially decreasing or increasing function of $x$.
These solutions are necessary in analyzing the scattering problem
in the Dirac electron system partially covered by a bulk superconductor
when the interface between covered and uncovered regions
is located along the $y$-axis.~\cite{beenakker}
In terms of the wave number in the $x$-direction,
\begin{align}
      \label{eq:def-k_x}
  k_{x}^{\pm}
   = \sqrt{\left(\frac{\mu\pm{\rm i}Z(\epsilon)\Theta(\epsilon)}{v}\right)^2
           -k_{y}^{2}} ,
\end{align}
the solutions of Eq.~(\ref{eq:BdG}) are expressed as
\begin{align}
    \label{eq:Psi_solution}
 \Psi_{\pm}^{\zeta}(x)
 & = C
     \left[\begin{array}{c}
             \left(\frac{\epsilon\pm{\rm i}\Theta(\epsilon)}
                        {\Delta(\epsilon)}\right)^{\frac{1}{2}}
             \left(\begin{array}{c}
                     vk_{-}^{\zeta} \\
                     \mu\pm{\rm i}Z(\epsilon)\Theta(\epsilon)
                   \end{array}
             \right)
             \\
             \left(\frac{\epsilon\mp{\rm i}\Theta(\epsilon)}
                        {\Delta(\epsilon)}\right)^{\frac{1}{2}}
             \left(\begin{array}{c}
                     vk_{-}^{\zeta} \\
                     \mu\pm{\rm i}Z(\epsilon)\Theta(\epsilon)
                   \end{array}
             \right)
           \end{array}
     \right]
     {\rm e}^{{\rm i}\zeta k_{x}^{\pm}x} ,
\end{align}
where $C$ is the normalization constant,
$\zeta (= \pm)$ specifies the direction of propagation, and
\begin{align}
  k_{-}^{\zeta}
  = \zeta k_{x}^{\pm}-{\rm i}k_{y} .
\end{align}
Unless $Z(\epsilon)\Theta(\epsilon)$ is pure imaginary, $k_{x}^{\pm}$ contains
an imaginary part and hence $\Psi_{\pm}^{\zeta}(x)$ becomes an exponentially
increasing or decreasing function of $x$.
When applying this to the scattering problem,
we need to choose only evanescent solutions.

Below, we demonstrate that the eigenfunction~(\ref{eq:Psi_solution})
is reduced to that derived from the conventional model
under the condition of $\mu \gg \Gamma \gg \Delta_{0}$.
Let us focus on the strong coupling limit of $\Gamma \gg \Delta_{0}$.
Note that, in this limit, the effective pair potential
$\Delta(\epsilon)$ becomes identical to $\Delta_{0}$,
being independent of $\epsilon$,
as is evident in Eq.~(\ref{eq:def-Del_eff}).
Additionally, if $\mu$ is much greater than $\Gamma$,
Eq.~(\ref{eq:Psi_solution}) can be approximated as
\begin{align}
    \label{eq:Psi_solution-0}
 \Psi_{\pm0}^{\zeta}(x)
 & = C
     \left[\begin{array}{c}
             \left(\frac{\epsilon\pm{\rm i}\Theta_{0}(\epsilon)}
                        {\Delta_{0}}\right)^{\frac{1}{2}}
             \left(\begin{array}{c}
                     vk_{-}^{\zeta} \\
                     \mu
                   \end{array}
             \right)
             \\
             \left(\frac{\epsilon\mp{\rm i}\Theta_{0}(\epsilon)}
                        {\Delta_{0}}\right)^{\frac{1}{2}}
             \left(\begin{array}{c}
                     vk_{-}^{\zeta} \\
                     \mu
                   \end{array}
             \right)
           \end{array}
     \right]
     {\rm e}^{{\rm i}\zeta k_{x}^{\pm}x}
\end{align}
with $\Theta_{0}(\epsilon)=\sqrt{\Delta_{0}^{2}-\epsilon_{+}^{2}}$.
It turns out that Eq.~(\ref{eq:Psi_solution-0}) is equivalent to
the solution of the BdG equation for $H_{\rm BdG}^{0}$
with the energy-independent pair potential $\Delta_{0}$
in the case of $\mu \gg \Delta_{0}$.~\cite{comment3}
That is, under the condition of $\mu \gg \Gamma \gg \Delta_{0}$,
the behavior of quasiparticles described by $H_{\rm BdG}(\epsilon)$
is equivalent to that described by $H_{\rm BdD}^{0}$.
This statement is also supported by the fact that, in the limit of
$\mu \gg \Gamma \gg \Delta_{0}$, the density of states~(\ref{eq:Dos_gen})
is reduced to
\begin{align}
 D(\epsilon)
   = \frac{1}{\pi v^{2}}
     \frac{\mu\epsilon}{\sqrt{\epsilon^{2}-\Delta_{0}^{2}}} ,
\end{align}
which is equivalent to that obtained from $H_{\rm BdG}^{0}$.

The above argument accounts for the correspondence that,
in the strong coupling limit of $\Gamma \gg \Delta_{0}$,
the expression for the Josephson current derived on the basis of
$H_{\rm BdG}(\omega)$~\cite{comment1} reproduces
that of the conventional model,~\cite{titov}
where $\mu \gg \Delta_{0}$ is assumed from the outset.

\section{Charge Conservation}

The evanescent solutions obtained above exponentially decay
with increasing or decreasing $x$, implying the disappearance of
quasiparticle current.
This missing current should be transferred to
the bulk superconductor coupled with the Dirac electron system.
In this section, we derive the charge conservation law
for quasiparticle states from the BdG equation.~\cite{blonder}

Let us express the four components of $\Psi(x)$ as
\begin{align}
  \Psi(x)={}^{\rm t}\!\left(u_{\uparrow}(x), u_{\downarrow}(x),
                        v_{\downarrow}(x), v_{\uparrow}(x)\right) ,
\end{align}
where $u_{\sigma}(x)$ and $v_{\sigma}(x)$ respectively are the electron
and hole wave functions for spin $\sigma$.
In terms of the electron charge $e$, we define the quasiparticle
charge density $Q(x)$ as~\cite{blonder}
\begin{align}
 Q(x) = e\left(|u_{\uparrow}(x)|^{2}+|u_{\downarrow}(x)|^{2}
               -|v_{\downarrow}(x)|^{2}-|v_{\uparrow}(x)|^{2}\right) .
\end{align}
In the stationary state in which we are interested, it is obvious that
the time derivative of $Q(x)$ vanishes:
\begin{align}
    \label{eq:Q1}
 \partial_{t}Q(x) = 0 .
\end{align}
On the other hand, by noting that ${\rm i}\partial_{t}\Psi(x)$
can be identified with $\epsilon\Psi(x)$ in the stationary state
with energy $\epsilon$ and then using the BdG equation~(\ref{eq:BdG}),
we can show that the quasiparticle charge density $Q(x)$ and
quasiparticle current density $J_{Q}(x)$ in the $x$-direction satisfy
\begin{align}
    \label{eq:Q2}
 \partial_{t}Q(x)
 = \Lambda_{Q}(x) + \Lambda_{S}(x) -\partial_{x}J_{Q}(x) ,
\end{align}
where
\begin{align}
     \label{eq:Lam_Q}
  \Lambda_{Q}(x)
  & = -2{\rm Im}\left\{Z(\epsilon)\right\}\epsilon Q(x) ,
         \\
     \label{eq:Lam_C}
  \Lambda_{S}(x)
  & = 4e{\rm Re}\left\{\phi(\epsilon)\right\}
      {\rm Im}\left\{u_{\uparrow}(x)^{\ast}v_{\downarrow}(x)
                    + u_{\downarrow}(x)^{\ast}v_{\uparrow}(x)\right\} .
\end{align}
The current density is expressed as
\begin{align}
     \label{eq:J_Q}
  J_{Q}(x)
    = \Psi(x)^{\dagger}\hat{J}_{Q}\Psi(x)
\end{align}
with
\begin{align}
  \hat{J}_{Q}
    = ev
      \left[ \begin{array}{cc}
                \check{\sigma}_{x} & \check{0} \\
                \check{0} & \check{\sigma}_{x}
             \end{array}
      \right] ,
\end{align}
where $\check{\sigma}_{x}$ is the $x$-component of the Pauli matrices.
Combining Eqs.~(\ref{eq:Q1}) and (\ref{eq:Q2}),
we arrive at the charge conservation law
\begin{align}
    \label{eq:cc_law}
  \Lambda_{Q}(x) + \Lambda_{S}(x) -\partial_{x}J_{Q}(x) = 0
\end{align}
in the stationary state.
It is obvious from Eqs.~(\ref{eq:Lam_Q}) and (\ref{eq:Lam_C}) that
$\Lambda_{Q}$ and $\Lambda_{S}$ are drain terms describing
charge tunneling into the superconductor:
$\Lambda_{Q}$ represents the contribution of quasiparticle tunneling,
while $\Lambda_{S}$ represents that of pair tunneling.
It should be emphasized that the conventional model does not involve
the drain term $\Lambda_{Q}(x)$ due to quasiparticle tunneling,
indicating its inadequacy in describing the proximity effect.

Let us examine charge transfer processes between the Dirac electron system
and the superconductor on the basis of Eq.~(\ref{eq:cc_law}).
In the region of $\Delta_{0} > \epsilon \ge 0$,
$Z(\epsilon)$ has no imaginary part, resulting in $\Lambda_{Q} = 0$.
This indicates that the quasiparticle tunneling plays no role,
reflecting the fact that the quasiparticle density of states
in the superconductor vanishes in this energy region.
To contrast, $\phi(\epsilon)$ becomes pure imaginary
in the region of $\epsilon > \Delta_{0}$, resulting in $\Lambda_{S} = 0$.
This indicates that pair tunneling plays no role.
To gain further insight into charge transfer processes,
we evaluate $\Lambda_{Q}$ and $\Lambda_{S}$ by substituting
the wave function given in Eq.~(\ref{eq:Psi_solution})
into Eqs.~(\ref{eq:Lam_Q}) and (\ref{eq:Lam_C}),
and confirm that the charge conservation law~(\ref{eq:cc_law})
actually holds (see Appendix B).
An important finding is that $\Lambda_{S} = 0$ in the region of
$\Delta_{0} > \epsilon > \epsilon_{\rm g}$.
Our result is summarized as
\begin{align}
  \Lambda_{S}(x) -\partial_{x}J_{Q}(x) = 0
\end{align}
for $\epsilon_{\rm g} > \epsilon \ge 0$,
\begin{align}
  \partial_{x}J_{Q}(x) = 0
\end{align}
for $\Delta_{0} > \epsilon > \epsilon_{\rm g}$, and
\begin{align}
  \Lambda_{Q}(x) -\partial_{x}J_{Q}(x) = 0
\end{align}
for $\epsilon > \Delta_{0}$.

It is worth pointing out that quasiparticle states in the region of
$\Delta_{0} > \epsilon > \epsilon_{\rm g}$ are
decoupled from the superconductor
in the sense that the corresponding quasiparticle current
is conserved within the Dirac electron system.
Supposing that the Dirac electron system is partially covered by
a bulk superconductor, let us consider the electron transport
from the uncovered region to the superconductor.
Note that electrons inevitably pass through the covered region
in the transport process.
We expect that quasiparticle states in the covered region with
$\Delta_{0} > \epsilon > \epsilon_{\rm g}$
do not contribute to the electron transport
since they are decoupled from the superconductor.
In contrast, quasiparticle states with $\epsilon_{\rm g} > \epsilon \ge 0$
do contribute to that
through the pair tunneling into superconducting condensate.
It is of interest
whether such contrasting behaviors can be experimentally observed.

\section{Summary}

We have studied the proximity effect on a two-dimensional massless
Dirac electron system in planar contact with a bulk superconductor,
starting from an appropriate microscopic model that explicitly
takes account of the coupling of Dirac electrons to the superconductor.
Integrating out the electron degrees of freedom in the superconductor,
we have derived a general proximity model for Dirac electrons.
The resulting effective model takes account of the proximity effect in terms of
the energy-dependent pair potential and renormalization term,
and is applicable regardless of the strength of coupling of
Dirac electrons to the superconductor.
From the analysis of the density of states, the quasiparticle wave function,
and the charge conservation of Dirac electrons,
it is shown that the effective model reveals several characteristic features
of the proximity effect, which cannot be captured by the conventional model,
implying its advantage over the conventional model.

Finally, it is worth mentioning that
the approach developed in this paper can be straightforwardly applied to
the hybrid system of multilayer graphene in planar contact 
with a bulk superconductor.~\cite{takane2,ludwig,takane3}

\section*{Acknowledgment}

This work is partially supported by a Grant-in-Aid for Scientific Research
(C) (No. 24540375).

\appendix

\section{Derivation of $S_{\Sigma}$}

To obtain the expression for $S_{\Sigma}$, we perform the integration over
$\psi_{j\sigma}$ and $\psi_{j\sigma}^{\dagger}$ in Eq.~(\ref{eq:def-Sigma})
following Ref.~\citen{affleck}.
It is convenient to rewrite the electron field as
\begin{align}
 \psi_{j\sigma}(\omega)
 = \frac{2}{\pi}\int_{0}^{\pi}{\rm d}q \sin(qj)\psi_{\sigma}(q,\omega) .
\end{align}
The substitution of this into the expression for $S_{\rm S}$ yields
\begin{align}
       \label{action_S_mod0}
   S_{\rm S}
 & = T\sum_{\omega} \frac{2}{\pi}\int_{0}^{\pi}{\rm d}q
     \left(\psi_{\uparrow}^{\dagger}(q,\omega),
           \psi_{\downarrow}(q,-\omega)\right)
       \nonumber \\
 & \hspace{-5mm}
     \times
     \left( \begin{array}{cc}
              -{\rm i}\omega+\xi_{q} & \Delta_{0} \\
              \Delta_{0} & -{\rm i}\omega-\xi_{q}
            \end{array}
     \right)
     \left( \begin{array}{c}
              \psi_{\uparrow}(q,\omega) \\
              \psi_{\downarrow}^{\dagger}(q,-\omega)
            \end{array}
     \right) ,
\end{align}
where $\xi_{q}=-2t\cos q-\mu_{\rm S}(\mib{k}_{\|})$.
We can simplify Eq.~(\ref{action_S_mod0})
in terms of the Bogoliubov transformation:
\begin{align}
   \left( \begin{array}{c}
            \psi_{\uparrow}(q,\omega) \\
            \psi_{\downarrow}^{\dagger}(q,-\omega)
          \end{array}
   \right)
   = \left( \begin{array}{cc}
              u_{q} & -v_{q} \\
              v_{q} & u_{q}
            \end{array}
     \right)
     \left( \begin{array}{c}
              \varphi_{+}(q,\omega) \\
              \varphi_{-}^{\dagger}(q,-\omega)
            \end{array}
     \right) ,
\end{align}
where
\begin{align}
  u_{q} & =\frac{1}{\sqrt{2}}\sqrt{1+\frac{\xi_{q}}{E_{q}}} ,
          \\
  v_{q} & =\frac{1}{\sqrt{2}}\sqrt{1-\frac{\xi_{q}}{E_{q}}}
\end{align}
with $E_{q}=\sqrt{\xi_{q}^{2}+\Delta_{0}^{2}}$.
Consequently, $S_{\rm S}$ is reduced to
\begin{align}
       \label{action_S_mod1}
   S_{\rm S}
 & = T\sum_{\omega} \frac{2}{\pi}\int_{0}^{\pi}{\rm d}q
     \sum_{\tau=\pm}
     \varphi_{\tau}^{\dagger}(q,\omega)
     \left(-{\rm i}\omega+E_{q}\right)
     \varphi_{\tau}(q,\omega) .
\end{align}
In terms of $\varphi_{\tau}$ and $\varphi_{\tau}^{\dagger}$,
the coupling term is rewritten as
\begin{align}
      \label{action_T_mod1}
  S_{\rm T}
 & = T\sum_{\omega}\frac{2}{\pi}\int_{0}^{\pi}{\rm d}q (-\gamma)\sin q
       \nonumber \\
 &  \times
    \Bigl[ \left(u_{q}\varphi_{+}^{\dagger}(q,\omega)
                 -v_{q}\varphi_{-}(q,-\omega)\right)
           \psi_{\rm D\uparrow}(\omega) + {\rm H.c.}
       \nonumber \\
 &  \hspace{2mm}
         + \left(v_{q}\varphi_{+}(q,\omega)
                 +u_{q}\varphi_{-}^{\dagger}(q,-\omega)\right)
           \psi_{\rm D\downarrow}(-\omega) + {\rm H.c.}
    \Bigr] .
\end{align}
We substitute Eqs.~(\ref{action_S_mod1}) and (\ref{action_T_mod1})
into Eq.~(\ref{eq:def-Sigma}) and replace the integration variables as
\begin{align}
  \prod_{j,\sigma,\omega}
  D\psi_{j\sigma}(\omega)D\psi_{j\sigma}^{\dagger}(\omega)
  \to
  \prod_{q,\tau,\omega}
  D\varphi_{\tau}(q,\omega)D\varphi_{\tau}^{\dagger}(q,\omega) .
\end{align}
We can obtain $S_{\Sigma}$ by integrating out
$\varphi_{\tau}$ and $\varphi_{\tau}^{\dagger}$.
It should be emphasized that this integration can be performed exactly
without relying on a perturbative treatment with respect to $S_{\rm T}$.
The result is
\begin{align}
   S_{\Sigma}
 & = T\sum_{\omega} \gamma^{2}\frac{2}{\pi}\int_{0}^{\pi}{\rm d}q \sin^{2}q
       \nonumber \\
 &   \hspace{0mm}
     \times
     \Bigg[ \left( \frac{v_{q}^{2}}{{\rm i}\omega+E_{q}}
                  -\frac{u_{q}^{2}}{-{\rm i}\omega+E_{q}}\right)
            \sum_{\sigma=\uparrow,\downarrow}
            \psi_{\rm D\sigma}^{\dagger}(\omega)\psi_{\rm D\sigma}(\omega)
       \nonumber \\
 &   \hspace{6mm}
          + \frac{2u_{q}v_{q}E_{q}}{\omega^{2}+E_{q}^{2}}
            \big(\psi_{\rm D\downarrow}(-\omega)\psi_{\rm D\uparrow}(\omega)
                  + {\rm H.c.}
            \big)
     \Bigg] .
\end{align}
Finally, we carry out the integration over $q$.
Since $2t$ is the largest energy scale of the system under consideration,
it is natural to assume that
$2t \gg \mu_{\rm S}(\mib{k}_{\|}), \Delta_{0}, |\omega|$.
We then arrive at
\begin{align}
   S_{\Sigma}
 & = T\sum_{\omega}
     \Bigg[ \left(\frac{\Gamma\mu_{\rm S}(\mib{k}_{\|})}{2t}
                        -\chi\frac{{\rm i}\Gamma\omega}{\Omega(\omega)}
            \right)
            \sum_{\sigma=\uparrow,\downarrow}
            \psi_{\rm D\sigma}^{\dagger}(\omega)\psi_{\rm D\sigma}(\omega)
       \nonumber \\
 &   \hspace{13mm}
          + \chi\frac{\Gamma\Delta_{0}}{\Omega(\omega)}
            \big(\psi_{\rm D\downarrow}(-\omega)\psi_{\rm D\uparrow}(\omega)
                  + {\rm H.c.}
            \big)
     \Bigg] ,
\end{align}
where $\Gamma = \gamma^{2}/t$,
$\Omega(\omega)=\sqrt{\omega^{2}+\Delta_{0}^{2}}$, and
\begin{align}
   \chi
   = 1-\frac{1}{2}\left(\frac{\mu_{\rm S}(\mib{k}_{\|})}{2t}
                  \right)^{2} .
\end{align}
This expression is equivalent to Eq.~(\ref{eq:Sigma_prox}).

\section{Check of Charge Conservation}

In this Appendix, we check that the charge conservation law~(\ref{eq:cc_law})
actually holds for the quasiparticle wavefunction $\Psi_{\pm}^{\zeta}(x)$
obtained in Sect. 4.
We examine only the case of $\zeta = +$,
for which  $\Psi_{\pm}^{+}$ is written as
\begin{align}
    \label{eq:Psi_pm^+}
 \Psi_{\pm}^{+}(x)
   = C
     \left[\begin{array}{c}
             \left(\frac{\epsilon\pm{\rm i}\Theta(\epsilon)}
                        {\Delta(\epsilon)}\right)^{\frac{1}{2}}
             \left(\begin{array}{c}
                     vk_{-}^{+} \\
                     \mu\pm{\rm i}Z(\epsilon)\Theta(\epsilon)
                   \end{array}
             \right)
             \\
             \left(\frac{\epsilon\mp{\rm i}\Theta(\epsilon)}
                        {\Delta(\epsilon)}\right)^{\frac{1}{2}}
             \left(\begin{array}{c}
                     vk_{-}^{+} \\
                     \mu\pm{\rm i}Z(\epsilon)\Theta(\epsilon)
                   \end{array}
             \right)
           \end{array}
     \right]
     {\rm e}^{{\rm i}k_{x}^{\pm}x} .
\end{align}
We separately treat the three energy regions
$\epsilon_{\rm g} > \epsilon \ge 0$,
$\Delta_{0} > \epsilon > \epsilon_{\rm g}$, and $\epsilon > \Delta_{0}$.
Remember that, in the first two cases, $\Lambda_{Q} = 0$
since ${\rm Im}\{Z(\epsilon)\}=0$ for $\Delta_{0} > \epsilon \ge 0$,
while $\Lambda_{S} = 0$ in the last case
since ${\rm Re}\{\phi(\epsilon)\} = 0$ for $\epsilon > \Delta_{0}$.

In the subgap region of $\epsilon_{\rm g} > \epsilon \ge 0$,
both $Z(\epsilon)$ and $\Theta(\epsilon)$ are real numbers.
Thus, Eq.~(\ref{eq:def-k_x}) indicates that
$k_{x}^{\pm}$ has an imaginary part, so we set
\begin{align}
    \label{eq:def-kappa}
  k_{x}^{\pm} = k_{0}^{\pm} + {\rm i}\kappa^{\pm} .
\end{align}
If $\kappa^{\pm}$ is positive (negative), $\Psi_{\pm}^{+}(x)$
is an exponentially decreasing (increasing) function of $x$.
From Eqs.~(\ref{eq:def-k_x}) and (\ref{eq:def-kappa}), we can show that
\begin{align}
     \label{eq:kappa_1}
  \kappa^{\pm}
  = \pm \frac{\mu Z(\epsilon)\Theta(\epsilon)}
             {v^{2}k_{0}^{\pm}} .
\end{align}
Since $\Lambda_{Q}=0$ in this case, we obtain $\Lambda_{S}$ and $J_{Q}$
for $\Psi_{\pm}^{+}(x)$ to check the charge conservation law.
Substituting Eq.~(\ref{eq:Psi_pm^+}) into Eqs.~(\ref{eq:Lam_C})
and (\ref{eq:J_Q}) and then using Eq.~(\ref{eq:def-k_x}),
we readily find that
\begin{align}
   \Lambda_{S}
 & = 8C^{2}\left[\mp Z(\epsilon)\Theta(\epsilon)\right]
     \left[\mu^{2}+v^{2}\kappa^{\pm}(\kappa^{\pm}-k_{y})
     \right] {\rm e}^{-2\kappa^{\pm}x} ,
        \\
   J_{Q}
 & = 4C^{2}v^{2}
     \left[k_{0}^{\pm}\mu
           \pm(\kappa^{\pm}-k_{y})Z(\epsilon)\Theta(\epsilon)
     \right] {\rm e}^{-2\kappa^{\pm}x} .
\end{align}
Modifying the expression for $\Lambda_{S}$ using Eq.~(\ref{eq:kappa_1}),
we confirm the following charge conservation law:
\begin{align}
  \Lambda_{S}(x) -\partial_{x}J_{Q}(x) = 0 .
\end{align}

In the region of $\Delta_{0} > \epsilon > \epsilon_{\rm g}$,
again $\Lambda_{Q}=0$ and $Z(\epsilon)$ is a real number.
However, $\Theta(\epsilon)$ becomes a pure imaginary number as
$\Theta(\epsilon)\equiv-{\rm i}\sqrt{\epsilon^{2}-\Delta(\epsilon)^{2}}$.
The substitution of Eq.~(\ref{eq:Psi_pm^+}) into Eq.~(\ref{eq:Lam_C})
straightforwardly yields $\Lambda_{S}=0$.
Correspondingly, $J_{Q}$ does not depend on $x$
as $k_{x}^{\pm}$ has no imaginary part in this region.
Taking these into account, we find that
\begin{align}
  \partial_{x}J_{Q} = 0
\end{align}
holds.

In the last case of $\epsilon > \Delta_{0}$,
both $Z(\epsilon)$ and $\Theta(\epsilon)$
contain real and imaginary parts.
Hence, $k_{x}^{\pm}$ has an imaginary part and is expressed
in the form of Eq.~(\ref{eq:def-kappa}).
It is convenient to decompose ${\rm i}Z(\epsilon)\Theta(\epsilon)$
into real and imaginary parts as
\begin{align}
     \label{eq:alpha-beta}
  {\rm i}Z(\epsilon)\Theta(\epsilon) = \alpha + {\rm i}\beta ,
\end{align}
in terms of which $\kappa^{\pm}$ is expressed as
\begin{align}
     \label{eq:kappa_2}
  \kappa^{\pm} = \pm\frac{(\mu\pm\alpha)\beta}{v^{2}k_{0}^{\pm}} .
\end{align}
Since $\Lambda_{S}=0$ when $\epsilon > \Delta_{0}$,
we obtain $\Lambda_{Q}$ and $J_{Q}$ for $\Psi_{\pm}^{+}(x)$.
Substituting Eq.~(\ref{eq:Psi_pm^+}) into Eqs.~(\ref{eq:Lam_Q}),
and (\ref{eq:J_Q}), we find after calculations using
Eqs.~(\ref{eq:def-k_x}) and (\ref{eq:alpha-beta}) that
\begin{align}
   \Lambda_{Q}
 & = -C^{2}\frac{\left|\epsilon\pm{\rm i}\Theta(\epsilon)\right|
                -\left|\epsilon\mp{\rm i}\Theta(\epsilon)\right|}
                {\left|\Delta(\epsilon)\right|}
     \frac{4\Gamma\epsilon}{{\rm i}\Theta(\epsilon)}
    \nonumber \\
 & \hspace{5mm}
     \times
     \left[(\mu\pm\alpha)^{2} + v^{2}\kappa^{\pm}(\kappa^{\pm}-k_{y})
     \right] {\rm e}^{-2\kappa^{\pm}x} ,
        \\
   J_{Q}
 & = 2C^{2}v^{2}\frac{\left|\epsilon\pm{\rm i}\Theta(\epsilon)\right|
                     +\left|\epsilon\mp{\rm i}\Theta(\epsilon)\right|}
                {\left|\Delta(\epsilon)\right|}
    \nonumber \\
 & \hspace{5mm}
     \times
     \left[k_{0}^{\pm}(\mu\pm\alpha)\pm\beta(\kappa^{\pm}-k_{y})
     \right] {\rm e}^{-2\kappa^{\pm}x} .
\end{align}
Modifying the expression for $\Lambda_{Q}$
using Eq.~(\ref{eq:kappa_2}) and the identity
\begin{align}
 & \big(\left|\epsilon\pm{\rm i}\Theta(\epsilon)\right|
       -\left|\epsilon\mp{\rm i}\Theta(\epsilon)\right|\big)\epsilon
     \nonumber \\
 &
 = \pm\big(\left|\epsilon\pm{\rm i}\Theta(\epsilon)\right|
          +\left|\epsilon\mp{\rm i}\Theta(\epsilon)\right|\big)
      \frac{{\rm i}\Theta(\epsilon)\beta}{\Gamma} ,
\end{align}
we can show that the charge conservation law, i.e.,
\begin{align}
  \Lambda_{Q}(x) -\partial_{x}J_{Q}(x) = 0 ,
\end{align}
actually holds.


\begin{thebibliography}{99}
\bibitem{comment0} Once the $T$-dependence of $\Delta_{0}$ is determined
by an appropriate gap equation based on BCS theory,
we can apply this model at arbitrary temperatures.
\bibitem{comment1} Equation~(83) in Ref.~\citen{takane2} represents
the Josephson current at $T = 0$ for monolayer and bilayer graphenes.
For monolayer graphene with $j=1$, this is reduced to Eq.~(19) in
Ref.~\citen{titov} if the following limiting procedure is performed:
$\Gamma_{1} \gg \Delta_{0}$ and $U \to \infty$.
The transmission probability $\tau_{n}$ in Ref.~\citen{titov} corresponds to
$2/\Phi_{1}$ in Ref.~\citen{takane2}.
\bibitem{comment2} Monolayer graphene has two different energy valleys,
and an electron in one valley couples to a hole in the other valley
under the superconducting proximity effect.
In contrast, the surface states of a typical strong topological insulator
have only one valley in which both an electron and a hole are present.
This difference plays no role in the argument given below
as long as the system is in the clean limit.
\bibitem{comment3} Strictly speaking, the $\epsilon$-dependences of
$k_{x}^{\pm}$ are different in the two cases.
However, this difference can be neglected as long as $\mu \gg \Delta_{0}$.
\end{thebibliography}
\end{document}